\documentclass[preprint,12pt,authoryear]{elsarticle}

\usepackage{natbib}
\usepackage{longtable}
\usepackage{multirow}

\usepackage{graphics}

\usepackage{amssymb}
\journal {New Astronomy}
\begin{document}

\begin{frontmatter}



\title{CCD Washington photometry of four poorly studied open clusters in the two inner quadrants of the 
galactic plane}



\author[oac]{N. Marcionni\corref{cor1}}
\ead{nahuel@oac.uncor.edu}
\author[oac,conicet]{J. J. Clari\'a \fnref{fn1}}
\ead{claria@oac.uncor.edu}
\author[oac,conicet]{M. C. Parisi}
\ead{celeste@oac.uncor.edu}
\author[oac,conicet]{T. Palma\fnref{fn1}}
\ead{tali@oac.uncor.edu}
\author[oac]{M. Oddone}
\ead{mao@oac.uncor.edu}
\author[oac,conicet]{A. V. Ahumada\fnref{fn1}}
\ead{andrea@oac.uncor.edu}
\address[oac]{Observatorio Astron\'omico, Universidad Nacional de C\'ordoba,
Laprida 854, C\'ordoba, Argentina}
\address[conicet]{Consejo Nacional de Investigaciones Cient\'{\i}ficas y T\'ecnicas, CONICET, Argentina}

\cortext[cor1]{Corresponding author}
\fntext[fn1]{Visiting Astronomer, CTIO.}

\begin{abstract}
Complementing our Washington photometric studies on Galactic open clusters (OCs), 
we now focus on four poorly studied OCs located in the first and fourth Galactic 
quadrants, namely BH\,84, NGC\,5381, BH\,211 and Czernik\,37. We have obtained 
CCD photometry in the Washington system $C$ and $T_1$ passbands down 
to $T_1$ $\sim$ 18.5 magnitudes for these four clusters. Their positions and sizes 
were determined using the stellar density radial profiles. We derived reddening, 
distance, age and metallicity of the clusters from extracted $(C-T_1,T_1)$ color-magnitude 
diagrams (CMDs), using theoretical isochrones computed for the Washington 
system. There are no previous photometric data in the optical band for BH\,84, NGC\,
5381 and BH\,211. The CMDs of the observed clusters show relatively well 
defined main sequences, except for Czernik\,37, wherein significant differential 
reddening seems to be present. The red giant clump is clearly seen only in BH\,211. 
For this cluster, we estimated the age in (1000$^{+260}_{-200}$) Myr, assuming a metallicity 
of $Z$ = 0.019. BH\,84 was found to be much older than it was previously believed, while 
NGC\,5381 happened to be much younger than previously reported. The heliocentric distances 
to these clusters are found to range between 1.4 and 3.4 kpc. BH\,84 appears to be located 
at the solar galactocentric distance, while NGC\,5381, BH\,211 and Czernik\,37 are 
situated inside the solar ring.

\end{abstract}


\end{frontmatter}

\section{Introduction}
\label{}
Galactic open clusters (OCs) have long been considered excellent targets not only 
to probe the Galactic disc properties \citep*{l82,f95,pietal06,bietal06} but also
to trace its chemical evolution (see, e.g., Chen et al., 2003 and references 
therein). Because it is relatively simple to estimate ages, distances and metallicities 
of OCs fairly accurately, their basic parameters constitute excellent tracers 
to the structure and chemical evolution of the Galactic disc. The proximity of most 
OCs to the Galactic plane, however, usually restricts this analysis to the most 
populous ones and/or to those located within a few kpc from the Sun (Bonatto et al., 2006). 
Although there are at present estimates of a total of about $25\times10^3$ OCs in the Milky Way 
\citep{poetal10}, there is not yet an estimation of fundamental parameters such as 
reddening, distance and age for nearly 30\% of the $\sim$ 2200 catalogued Galactic OCs 
\citep*{dietal02, bietal03,duetal03}.

The present work is part of a current project of photometric observation of Galactic 
OCs in the Washington system that is being developed at the Observatorio 
Astron\'omico de la Universidad Nacional de C\'ordoba (Argentina). This project aims at 
determining the fundamental parameters or at refining the quality of observationally 
determined properties for some unstudied or poorly studied OCs, located in different 
regions of the Milky Way. Washington photometry has proved to be a valuable tool to determine 
the fundamental parameters of OCs since information on cluster membership, reddening, 
distance, age and metallicity is obtained through the analysis of the 
$(C-T_1,T_1)$ color-magnitude diagram (CMD). We have already reported results 
based on Washington CCD $CT_1$ photometric data on several young (e.g., Piatti 
et al. 2003a), intermediate-age (e.g., Clari\'a et al., 2007) and old Galactic 
OCs (e.g., Piatti et al. 2004). These studies have contributed not only to the 
individual characterization of these stellar systems but also to the global 
understanding of some properties of the Galactic disc (e.g., Parisi et al. 2005).

In this study we provide new high-quality photometric CCD data obtained with the 
Washington system $C$ and $T_1$ passbands down to $T_1$ $\sim$ 18.5 magnitudes 
in the fields of four faint, poorly studied OCs, namely BH\,84, NGC\,5381, BH\,211 
and Czernik\,37. The equatorial and Galactic coordinates of the cluster 
centers taken from the WEBDA Open Cluster Database \citep{me05} are listed in 
Table 1, together with the angular sizes given by \citet{ah03}. The selected 
clusters are located in the first and fourth Galactic quadrants (280$^\circ$ $<$ 
{\it l} $<$ 3$^\circ$) near the Galactic plane ($\mid$b$\mid$ $\leq$ 3$^\circ$). As far 
as we know, no previous photometric data in the optical band exist for BH\,84, 
NGC\,5381 and BH\,211. The four selected clusters have been examined by 
\citet{buetal11} and \citet{ta08,ta11} using Two-Micron All-Sky Survey (2MASS) 
data. Some preliminary results about BH\,84, NGC\,5381 and BH\,211 are presented in
\citet{maetal13}. A brief description of these objects as well as a summary of previous 
results for the fields under investigacion is given below:

{\it BH\,84}. First recognized as an OC by \citet{vh75}, this object (IAU designation 
C0959-579) seems to be a detached, relatively poor and faint OC in the Carina 
constellation (Fig. 1). It shows the typical morphology of a Trumpler class II-1p 
cluster, which is characterized by a slight concentration of member stars of 
similar brightness bf and relatively small population. The only observational data for this object 
are those given in the 2MASS catalog and discussed by \citet{buetal11}. These 
authors derived a reddening $E(B-V)$ = 0.60 and suggest that BH\,84 is a young 
cluster ($\sim$ 18 Myr), located at a heliocentric distance $d$ = (2.92 $\pm$ 
0.19) kpc.

{\it NGC\,5381}. This is a cluster in Centaurus, also designated as BH\,156 
\citep{vh75}. \citet{ah03} refer to this object as belonging to Trumpler class 
II-2m, i.e., a moderately rich, detached cluster with little central concentration 
and medium-range bright stars (Fig. 1). According to these authors, NGC\,5381 
has a comparatively large angular diameter of 11$'$. A search for variable stars 
in the cluster field was carried out by \citet{pietal97}. Using 2MASS data, 
\citet{ta11} suggests that NGC\,5381, slightly reddened by $E(B-V)$ = 0.06, is an 
intermediate-age cluster ($\sim$ 1.6 Gyr) located at 1.2 kpc from the Sun.

{\it BH\,211}. This object (C1658-410) appears to be somewhat elongated in the East-West 
direction (Fig. 1). BH\,211 is a detached, moderately poor and relatively 
faint group of stars, first recognized as an OC in Scorpius by \citet{vh75}. It is 
a small-sized OC situated very near the Galactic center direction, practically on 
the Galactic plane (Table 1). The only observational data-set for this object is the one 
given in the 2MASS catalog and discussed by \citet{buetal11} who found the following 
results:  $E(B-V)$ = 0.48, $d$ = (1.38 $\pm$ 0.09) kpc and $\sim$ 1.6 Gyr.

{\it Czernik\,37}. Also known as BH\,253 \citep{vh75}, this is a relatively faint 
cluster (C1750-273) first recognized in Sagittarius by \citet{cz66}. As indicated 
by its Trumpler class (II-1m), it shows a slight central concentration but can 
be identified by its relatively dense population compared to that of the field 
stars (Fig. 1). This cluster is projected on to the central bulge of the Galaxy, 
only 2$^\circ$ from the Galactic center direction. \citet{caetal05} presented CCD
$BVI$ photometry in the field of Czernik\,37. Although they conclude that this may 
be a sparse but real cluster superimposed on the Galactic bulge population, they 
do not provide its physical parameters. Using 2MASS data, \citet{ta08} derived a 
heliocentric distance of 1.7 kpc, $E(B-V)$ = 1.03 and an age of 0.6 Gyr.

The layout of this paper is as follows. Section 2 provides details on our 
observations and the data reduction procedure. In Section 3 we determine the cluster 
centers and the stellar density radial profiles. Section 4 deals with the determination 
of cluster fundamental parameters through the fitting of theoretical isochrones. A 
brief description of the results, including a comparison with previous findings, is 
presented in Section 5, while the final conclusions are summarized in Section 6.

\section{Observations and reductions}
The observations of the selected clusters were carried out with the Cerro Tololo 
Inter-American Observatory (CTIO, Chile) 0.9 m telescope, during the nights of 2008 May 9-11, 
with a 2048$\times$2048 pixel Tektronix CCD. The scale on the chip is 0.396" pixel$^{-1}$ (focal 
ratio f/13.5) yielding a visual field of 13.6'$\times$13.6'. We controlled the CCD through 
the CTIO ARCON 3.3 data acquisition system in the standard quad amplifier mode,  with a mean 
gain and readout noise of 1.5 e$^-$/ADU and 4.2 e$^-$, respectively. The filters used were the
Washington system $C$ \citep{ca76} and Kron-Cousins $R_{KC}$. The latter has a significantly higher 
through-put as compared with the standard Washington $T_1$ filter but $R_{KC}$ magnitudes can be 
accurately transformed to yield $T_1$ magnitudes \citep{ge96}. From here onwards, we will use indistinctly 
the words $R_{KC}$ or $T_1$. Typically 20 standard stars taken from the list of \citet{ge96}, 
covering a wide range in color, were nightly observed. Table 2 shows the log of the observations 
with dates, filters, exposure times and airmasses. In addition, a series of 10 bias and five 
dome and sky flat-field exposures per filter were obtained nightly. The weather conditions kept 
very stable at CTIO, with a typical seeing of 1.0"-1.2", although some images have slightly larger 
full-widths at half maximum (FWHMs) due to temperature changes of up to 2 $^\circ$C. Fig. 1 shows 
schematic finding charts of the observed cluster fields built with all the measured stars in the $T_1$ band.

The $CT_1$ images were reduced at the Observatorio Astron\'omico de la Universidad Nacional de 
C\'ordoba (Argentina) with IRAF\footnote{IRAF is distributed by the National Optical Astronomy 
Observatories, which is operated by the Association of Universities for Research in Astronomy, Inc., 
under contract with the National Science Foundation}, using the QUADPROC package. After applying 
the overscan-bias subtraction for the four amplifiers independently, we carried out flat-field corrections 
using a combined sky-flat frame, which was previously checked for nonuniform illumination pattern with 
the averaged dome-flat frame. Then, we did aperture photometry for the standard star fields 
using the PHOT task within DAOPHOT II \citep{st91}. The relationships between instrumental 
and standard magnitudes were obtained by fitting the  equations:

\begin{equation}
c = a_1 + T_1 + (C-T_1) + a_2\times X_C + a_3\times (C-T_1),
\end{equation}

\begin{equation}
r = b_1 + T_1 + b_2\times X_{T_1} + b_3 \times (C-T_1),
\end{equation}

\noindent where $X$ represents the effective airmass and capital and lowercase letters 
stand for standard and instrumental magnitudes, respectively. The  coefficients $a_i$ and 
$b_i$ ($i$ = 1, 2 and 3) were nightly derived through the IRAF routine FITPARAM. The resulting 
mean coefficients together with their errors are shown in Table 3; the typical rms errors of 
equations (1) and (2) are 0.017 and 0.015 mag, respectively, indicating the nights were 
of good  photometric quality.

Point spread function (PSF) photometry for the selected fields was performed using the 
stand-alone version of the DAOPHOT II package \citep{st94}, which provided us with $x$ and 
$y$ coordinates and instrumental $c$ and $r$ magnitudes for all the stars identified in each 
field. The PSFs were generated from two samples of 35-40 and $\sim$ 100 stars interactively selected.
For each frame, a quadratically varying PSF was derived by fitting the stars in the larger sample, 
once their neighbors were eliminted using a preliminary PSF obtained from the smaller star 
sample, which contained the brightest, least contaminated stars. We then used ALLSTAR program to 
apply the resulting PSF to the identified stellar objects and create a subtracted image, which was 
used to find and measure magnitudes of additional fainter stars. The PSF magnitudes were determined 
using as zero points the aperture magnitudes yielded by PHOT. This procedure was repeated three times  
on each frame. Next, we computed aperture corrections from the comparison of PSF and aperture  
magnitudes using the subtracted neighbor PSF star sample. The resulting aperture corrections were 
on average less than 0.02 mag (absolute value) for $c$ and $r$, respectively. Note that PSF stars 
are distributed throughout the whole CCD frame, so that variations of aperture corrections should 
be negligible. Finally, the standard magnitudes and colors for all the measured stars were computed by 
inverting equations (1) and (2), once positions and instrumental $c$ and $r$ magnitudes of stars in 
the same field were matched using Stetson's DAOMATCH and DAOMASTER programs.

Once we obtained the standard magnitudes and colors, we built a master table containing the average 
of $T_1$ and $C-T_1$, their errors $\sigma$$(T_1)$ and $\sigma$$(C-T_1)$ and the number of observations 
for each star, respectively. We derived magnitudes and colors for 1884, 2439, 979 and 1129 stars 
in the fields of BH\,84, NGC\,5381, BH\,211 and Czernik\,37, respectively. These 
values are provided in Tables 4-7. Magnitude and color errors are the standard 
deviation of the mean or the observed photometric errors for stars with only 
one measurement. Tables 4-7 are only partially presented here as guidance, 
regarding their form and content. The complete tables can be found on the on-line 
version of the journal.

Fig. 2 shows the behavior of the $T_1$ and $C-T_1$ photometric errors as a 
function of $T_1$ for stars measured in the field of NGC\,5381, the most populated
observed field. Note that $\sigma$$T_1$ and $\sigma$$(C-T_1)$ appear to be smaller 
than 0.04 magnitudes for stars brighter than $T_1$ $\sim$ 18.5 and 17.5 magnitudes, 
respectively. The bright stars having large associated errors are stars saturated 
in all frames, bits of stars or even failures  in the detector erroneously taken as stars by 
DAOPHOT. It can be observed in Section 4 that these stars do not affect at all the fundamental 
parameter determination of the observed clusters. Bearing in mind the behavior of the 
photometric errors with the $T_1$ magnitude for the observed stars in Fig. 2, we trust the accuracy of 
the morphology and position of the main clster features in the $(C-T_1,T_1)$ 
CMDs. Fig. 3 shows these CMDs for all the observed stars in the different fields. 
They appear to be contaminated by field stars, as should be expected since all the studied 
OCs are projected on to the Galactic center direction. As can be seen in Fig 3, 
the faintest $T_1$ magnitudes obtained for each cluster are such that they allow the 
mapping of most of the cluster main sequences (MSs). If there are still fainter stars in 
the clusters, then they practically do not add any information to the one already obtained. 
Cluster MSs appear as broad sequences of stars among crowded field features. 
On the other hand, a group of comparatively bright late-type stars in BH\,211 
seems to form the cluster red giant clump (RGC) centered at $T_1$ $\sim$ 
12.5 and $(C-T_1)$ $\sim$ 3.2 magnitudes, which indicates that we are possibly dealing 
with an intermediate-age OC.

\section{Determining the center and size of the clusters}
To determine the central position of the clusters on a firm basis, we applied 
a statistical method consisting in tracing the stellar density profiles 
projected on to the directions of the $x$ and $y$-axes. The coordinates of the 
clusters' centers and their estimated uncertainties were determined by 
fitting Gaussian distributions to the star counts in the $x$ and $y$ directions 
for each cluster. The fits of the Gaussians were performed using the NGAUSSFIT 
routine in the STSDAS/IRAF package. We adopted a single Gaussian and decided to 
fix the constant and the linear terms to the corresponding background levels and 
to zero, respectively. We used as variables the center of the Gaussian, its 
amplitude and its FWHM. After eliminating a couple 
of scattered points, the fitting procedure converged after one iteration on 
average. The stars projected along the $x$ and $y$ directions were counted 
within intervals of 50 pixels in BH\,84 and NGC\,5381 and within intervals of 
75 pixels in BH\,211 and Czernik\,37. We determined the central position of 
the clusters with a typical NGAUSSFIT standard deviation of $\pm$ 10 pixels 
($\sim$ 4") in all cases. The centers of the Gaussians were finally fixed at 
($X_C,Y_C$) = (969,1040), (940,1012), (964,1145) and (894,1105) pixels for 
BH\,84, NGC\,5381, BH\,211 and Czernik\,37, respectively. These values 
were adopted for the analysis that follows. Cluster centers are marked by a 
cross in Fig. 1.

We then built clusters' radial profiles, from which we estimated the cluster 
radii, generally used as an indicator of cluster dimensions. Cluster stellar 
density radial profiles are usually built by counting the number of stars 
distributed in concentric rings around the cluster center and normalizing the 
sum of stars in each ring to the unit area. Although this procedure allows 
us to stretch the radial profile to its utmost, until complete circles can  
no longer be traced in the observed field, we preferred to follow another method in 
order to move even further away from the cluster center. This method is based 
on counts of stars located in boxes of 50 pixels a side, distributed throughout 
the whole field of each cluster. Thus, the number of stars per unit area 
at a given $r$ can be directly calculated through the expression:

\begin{equation}
(n_{r+25} - n_{r-25})/[(m_{r+25}-m_{r-25})\times 50^{2}],
\end{equation}
                    
\noindent where $n_j$ and $m_j$ represent the number of counted stars and boxes 
included in a circle of radius $j$, respectively. Note that this procedure 
does not necessarily require a complete circle of radius $r$ within the 
observed field to estimate the mean stellar density at such distance. It is 
important to consider this fact since having a stellar density profile which 
extends far away from the cluster center allows us to estimate the background 
level with higher precision. This is necessary to derive the cluster radius 
($r_{cl}$), defined as the distance from the cluster center where the stellar 
density profile intersects the background level. It is also helpful to 
measure the FWHM of the stellar density profile, which plays an important 
role in the construction of the cluster CMDs. The resulting density profiles 
expressed as number of stars per unit area in arcmin$^2$ are presented in 
Fig. 4. The uncertainties estimated at various distances from the cluster 
centers follow Poisson statistics. The new equatorial coordinates derived for 
the clusters in this study are listed in Table 1, while columns 2-4 of Table 
8 list the radii at the FWHM ($r_{FWHM}$), the estimated radii which yield
the best enhanced cluster fiducial features ($r_{clean}$) and the cluster 
radii ($r_{cl})$. The different linear radii in parsecs were determined 
using the heliocentric distances derived in Section 4.

\section{Cluster fundamental parameter estimates}
CMDs covering different circular extractions around each cluster center were 
constructed, as shown in Figs. 5-8. The panels in the figures are arranged, 
from left to right and from top to bottom, in such a way that they exhibit the
variations in stellar population from the innermost to the outermost regions 
of the cluster fields. We started with the CMD for stars distributed within 
$r$ $<$ $r_{FWHM}$, followed by those of the cluster regions delimited by 
$r$ $<$ $r_{clean}$ and $r$ $<$ $r_{cl}$ and finally by the adopted field CMD. 
We used the CMDs corresponding to the stars within $r_{FWHM}$ as the cluster 
fiducial sequence references. Then, we varied the distance from the cluster 
centers, starting at $r_{FWHM}$. Next, we built different series of extracted 
CMDs. Finally, we chose those CMDs - one per cluster - which maximize the 
star cluster population and minimize the field star contamination in the 
CMDs. The main cluster features can be identified by inspecting the right top 
panels of Figs. 5-8. Note that the cluster MSs look well populated, particularly 
in NGC\,5381, all of them showing clear evidence of evolution. These MSs 
develop along $\sim$ 5 magnitudes in BH\,84 and Czernik\,37 and along $\sim$ 5-6 
magnitudes in NGC\,5381 and BH\,211. The hook at the MS turnoff of BH\,211's 
CMD and the RGC centered at $T_1$ $\sim$ 12.5 and $C-T_1$ $\sim$ 3.2 
magnitudes indicate that we are dealing with an intermediate-age OC. The 
width of the clusters' MSs is clearly not the result of photometric 
errors, since these hardly reach a tenth of magnitude at any $T_1$ level. 
Thus, such width could be caused by intrinsic effects (binarity, rotation, evolution, 
etc.), by differential reddening and/or by field star contamination. To derive 
the cluster fudamental parameters, we will use the CMDs with $r$ $<$ $r_{clean}.$ 

Although, as we mentioned above, the broadness of the clusters' MSs is certainly due 
to several effects, in the particular case of Czernik\,37, a clear variation of the 
interstellar reddenning across the cluster field seems to be present. In case this 
effect indeed existed in the remaining clusters, it is by far less evident. The 
lower limit estimated by \citet{b75} for clusters with differential reddening is 
$\Delta$$(B-V)$ = 0.11, which corresponds to $\Delta$$(C-T_1)$ = 0.22, if a value of 
1.97 for the $E(C-T_1)$/$E(B-V)$ ratio \citep{ge96} is adopted. From the right top 
panel of Fig. 8, we estimated $\Delta$$(C-T_1)$ $\sim$ 1.0 mag for Czernik\,37, a 
value which largely exceeds the limit established by \citet{b75}. The existence of 
differential reddening in Czernik\,37 makes the determination of its fundamental 
parameters very difficutl, particularly its reddening and its heliocentric distance. 
It is for this reason that the resulting parameters for Czernik\,37 present associated 
errors larger than  in the other clusters.

The widely used procedure of fitting theoretical isochrones to observed 
CMDs was employed to estimate the $E(C-T_1)$ color excess, the $T_1$-M$_{T_1}$ 
apparent distance modulus and the age and metallicity of the clusters. 
It is well known that the metallicity of a cluster plays an important 
role when its age is estimated from the fit of theoretical isochrones. 
Indeed, isochrones with the same age but with different metallicities 
can range from slightly to remarkably distinguishable, depending on 
their sensitivity to metal content. The $(C-T_1,T_1)$ CMD, for example, 
is nearly three times more metal sensitive than the $(V-I,V)$ CMD \citep{gs99}.
The distinction is particularly evident for the evolved phases of the 
RGC and the giant branch. As far as zero-age main sequences (ZAMSs) are 
concerned, they are often less affected by metallicity effects, and can even 
exhibit imperceptible variations for a specific metallicity range within 
the photometric errors. This is the case of Galactic OCs, therefore including the 
four clusters here studied. Since there are no previous available estimates 
of their metallicities, we followed the general rule of starting by the adoption 
of both solar ($Z$ = 0.019) and sub-solar ($Z$ = 0.008) values for the clusters' 
metal content.

As for the isochrone sets, we used those computed by the Padova group \citep{gietal02} 
in steps of $\Delta$log age = 0.05 dex. As shown in previous studies (e.g., 
Piatti et al., 2003b), these isochrones lead to results similar to those 
derived from the Geneve group's isochrones \citep{ls01}. We preferred to 
use the Padova group's isochrones because they reach  fainter magnitudes, thus 
allowing a better fit to the cluster fainter portions of the MSs. Then, we 
first selected ZAMSs with $Z$ = 0.019 and 0.008 ([Fe/H] = 0.0 and -0.4 dex) and 
fitted these ZAMSs to the cluster CMDs to derive color excesses and apparent 
distance moduli for each selected metallicity. Since the fits are performed 
through the lower envelope of the cluster's MSs, the presence of binaries 
practically does not affect the choice of the best isochrones. Note that 
when $Z$ = 0.019 is used in the fits, the resulting cluster reddenings and 
distances turn out to be slightly larger than the values obtained using Z = 0.008.
The increases in distance vary from 250 pc to 500 pc depending on the cluster,
while the increase in reddening is within error limits.
Then, using each of the derived 
[$E(C-T_1)$,$T_1$-M$_{T_1}$]$_{Z}$ sets, we performed isochrone fits. We 
repeated the fits for a larger number of isochrones covering appropriate age 
ranges according to each cluster. The brightest magnitude in the MS, the 
bluest point of the turnoff and the locus of the RGC, when visible, were 
used as reference points during the fits. Finally, we chose the best fit for 
each [$E(C-T_1)$,$T_1$-M$_{T_1}$]$_{Z}$ set and compared all the individual 
best fits to choose the one which best reproduced the cluster features. In 
all cases, the best fits were done only by eye and they were obtained 
using solar metallicity isochrones.
We would like to point out that the bright stars with very large 
associated errors in Fig. 2 correspond to objects located beyond NGC\,5381 
$r_{clean}$. For this reason, these stars do not affect the cluster 
parameter determination in NGC\,5381. Something similar occurs with the 
other three studied clusters.

Fig. 9 illustrates the results of our task, while Table 9 lists the estimated 
$E(B-V)$ color excesses, heliocentric distances, ages and metallicities of the 
clusters. Their errors were derived taking into account the broadness of 
the cluster MSs and, in the case of the cluster distances, the expression 
0.46$\times$[$\sigma$($V-M_{V}$) + 3.2$\times$$\sigma$$E(B-V)$]$\times$d, 
where $\sigma$$(V-M_{V})$ and $\sigma$$(E(B-V))$ represent the estimated errors in 
$V-M_{V}$ and $E(B-V)$, respectively (see, e.g., Clari\'a et al., 2007). The 
expressions $E(C-T_1)$ = 1.97$\times$$E(B-V)$ and M$_{T_1}$ = 
$T_1$ + 0.58$\times$$E(B-V)$ - (V-M$_{V}$) taken from \citet{ge96} were used 
to relate both color excesses and distance moduli. We also list in Table 9 
the Galactocentric rectangular coordinates $X$,$Y$,$Z$ and the Galactocentric 
distances $R_{GC}$ of the clusters, derived assuming the Sun's distance from 
the center of the Galaxy to be 8.5 kpc. The adopted reference system is 
centered on the Sun, with the $X$ and $Y$-axes lying on the Galactic plane and 
$Z$ perpendicular to the plane. $X$ points towards the Galactic center, being 
positive in the first and fourth quadrants; $Y$ points in the direction 
of the Galactic rotation, being positive in the first and second Galactic 
quadrants. $Z$ is positive towards the north Galactic pole.

\section{Results}
The 2MASS catalog has been widely used to determine the fundamental 
parameters of hundreds of OCs (see, e.g., Kronberger et al., 2006; Bonatto 
\& Bica 2007; Tadross 2011). The results obtained, however, do not always yield 
consistent results with those derived using optical data. As shown by 
\citet{dietal12}, the accuracy in determining the color excess $E(J-H)$ 
using only 2MASS is mainly limited by structural uncertainty in the 
MS and/or narrow range of magnitude sampling of the MS. From the 
observational point of view, the main reason for this discrepancy is the 
limiting magnitude, particularly for the oldest clusters, which reflects 
in the sampling of the MS, together with photometric errors. Note that 
2MASS photometric errors typically reach 0.10 magnitudes at $J$ $\leq$ 
16.2 and $H$ $\leq$ 15.0 \citep{sb02}, while in the optical bands 
($UBV$ or $CT_1$, for example), they are typically lower than 0.05 
magnitudes at $V$ $\leq$ 18.0 and $T_1$ $\leq$ 19.0. In the following 
subsections, we compare the current results with those obtained in previous 
studies using 2MASS.

\subsection{BH\,84}
The equatorial coordinates we derived for the center of BH\,84 differ only 
by 33" in declination from the WEBDA value (Table 1). The radial number 
density profile of BH\,84 is shown in Fig.4, from which we derive a 
radius of 3.6', slightly lower than the value reported by \citet{ah03}. 
The cluster MS is easily identifiable in Fig. 5, extending along $\sim$  
5 magnitudes and with clear evidence of some evolution. Outer Galactic 
disc stars remarkably contaminate the cluster MS, particularly its 
fainter half portion. The solar metallicity ZAMS fitted to the cluster CMD 
yields a reddening $E(C-T_1)$ = 1.25 $\pm$ 0.10, equivalent to $E(B-V)$ = 
0.63 $\pm$ 0.05, and a true distance modulus $(T_1)$$_{o}$-M$_{T_1}$ = 
12.64 $\pm$ 0.15, which implies a distance of (3.37 $\pm$ 0.48) kpc from 
the Sun. For $Z$ = 0.019, the best-fitting isochrone corresponds to an 
age of (560$^{+150}_{-110}$) Myr (Fig. 9). Therefore, BH\,84, situated at 142 pc 
below the Galactic plane and $\sim$ 8.58 kpc from the Galactic center (Table 9), 
is found to be a cluster only slightly younger than the Hyades. Both the 
reddening and the distance here derived show a reasonable agreement with the results 
from 2MASS data \citep{buetal11}. Surprisingly, however, the age we 
find is significantly larger than \citet{buetal11} estimate of 18 Myr. 
Although some of the 2MASS error must be due to the larger pixels, we believe 
the authors' 
gross underestimation of BH\,84 age is clearly the result 
of the impossibility to see the turnoff point in the $(J-K,J)$ diagram. 
Hence, \citet{buetal11} derived the cluster's age just by fitting the 
ZAMS. 

\subsection{NGC\,5381}
This appears to be a large cluster covering nearly the entire observed 
field (Fig.1). We derived for NGC\,5381 practically the same equatorial 
coordinates as those listed in WEBDA (Table 1). It is difficult, however, 
to estimate the cluster radius from the stellar density radial profile 
(Fig. 4), because it does not present a typical cluster-like structure. 
If the area for $r$ $>$ 1200 pixels is considered to be the ``star field 
area'', then NGC\,5381 seems to have a relatively small but conspicuous 
nucleus and a low-density extended corona (see also Fig. 1). We estimate 
the angular core and corona radii as $\sim$ 330 pixels ($\sim$ 2.2') 
and 870 pixels ($\sim$ 5.7'), respectively. Fig. 3 reveals a crowded 
broad sequence of stars that traces the cluster MS along $\sim$ 5-6 
magnitudes and with clear signs of evolution. This fact suggests 
that the age of the cluster is some hundred million years. A number of 
stars visible in Fig. 9 with $T_1$ magnitudes between 18 and 19 
and $(C-T_1)$ colors ranging from 0.5 up to 1.5 magnitudes are 
clearly background stars. No clump of red stars is visible. However, 
the analysis of 2MASS data by \citet{ta11} reveals that NGC\,5381 is 
an intermediate-age (1.6 Gyr) cluster, suffering low interstellar extinction. 
This is not confirmed by the current optical data. In fact, our best fit 
of the solar metallicity ZAMS in the $(C-T_1,T_1)$ CMD yields $E(C-T_1)$ 
= 0.90 $\pm$ 0.08, equivalent to $E(B-V)$ = 0.46 $\pm$ 0.04, and a 
heliocentric distance of (2.63 $\pm$ 0.40) kpc, while an age of about 
(250$^{+65}_{-50}$) Myr is derived from the best-fitting isochrone (Fig. 9). Therefore, 
NGC\,5381 now appears to be significantly younger than previously 
believed. The age difference between ours and \citet{ta11} may depend 
at least partially on the reddening difference. Our reddening estimate, 
$E(B-V)$ = 0.46, is quite larger than the value of $E(B-V)$ = 0.06 
reported by \citet{ta11}. Therefore, a smaller age value is obtained 
when a larger reddening value is adopted. NGC\,5381 lies at $\sim$ 100 
pc above the Galactic plane and $\sim$ 7.04 kpc from the Galactic 
center (Table 9).

\subsection{BH\,211}
The coordinates for the cluster center derived in the current study 
differ by $\sim$ 15" in right ascension and by only 3" in declination 
from the WEBDA values (Table 1). From the stellar density radial 
profile (Fig. 4), we determined the radius of BH\,211 to be 
$\sim$ 3.9', in very good agreement with the value reported by \citet{ah03}. 
The $(C-T_1,T_1)$ CMD obtained using all the measured stars in the cluster 
field is depicted in Fig. 3, wherein the main cluster features can be 
identified. What first calls our attention is the cluster MS, which 
looks well populated, has clear signs of evolution and develops along 
$\sim$ 5-6 magnitudes. It is relatively broad, especially in its lower 
envelope, partly due to field star contamination. Another interesting 
feature of the $(C-T_1,T_1)$ CMD is the presence of a number of good 
candidates for giant clump stars centered at $(C-T_1,T_1)$ = (3.2,12.5).
Since these stars lie within 400 pixels (2.6') from the cluster center, 
we may reasonably expect many of them to be cluster giants. The 
reddening and age we find, $E(B-V)$ = 0.61 $\pm$ 0.05 and $\sim$ 
(1000$^{+260}_{-200}$) Myr (Table 9), turn out to be somewhat larger 
and lower, respectively, than \citet{buetal11} estimates. In any case, 
there seems to be no doubt that BH\,211 is an intermediate-age cluster 
located at about (1.44 $\pm$ 0.21) kpc from the Sun, at scarcely $\sim$ 12 
pc out of the Galactic plane and $\sim$ 7.12 kpc from the Galactic center 
(Table 8).

\subsection{Czernik\,37}
The equatorial coordinates we derived for this cluster are different by 
15" in right ascension and by 36" in declination (Table 1) from the 
WEBDA values. The radial number density profile of Czernik\,37 is 
shown in Fig. 4, from which we derived a radius of 3.6', in good 
agreement with the value reported by \citet{ah03}. This cluster 
appears to be embedded in the dense stellar population towards the 
Galactic bulge so that their properties are difficult to be determined. 
Although the cluster CMD (Fig. 3) is profusely contaminated by field 
stars, there is an appearance of a broad MS with evidence of some 
evolution. As mentioned in Section 4, the broadness of the MS of 
Czernik\,37 is caused not only by field star contamination and other 
effects (binarity, rotation, photometric errors, ets.) but also by 
differential reddening. Such effect is probably due to the presence of 
irregularly distributed dark clouds, projected towards the cluster. Different 
isochrone fittings using $Z$ = 0.019 in Fig. 9 allow us to estimate the 
variation range of $E(C-T_1)$ as $\sim$ 1.0 mag. The solar metallicity ZAMS
which best fits the cluster CMD (Fig. 9) yields a reddening $E(C-T_1)$ = 2.90 
$\pm$ 0.50, equivalent to 
$E(B-V)$ = 1.47 $\pm$ 0.25, and a true distance modulus $(T_1)$$_{o}$-
M$_{T_1}$ = 10.80 $\pm$ 0.50. The isochrone corresponding to an age 
of (250$^{+100}_{-65}$) Myr reasonably reproduces the main cluster features 
in Fig. 9. These results place Czernik\,37 at a distance of (1.44 
$\pm$ 0.86) kpc from the Sun and $\sim$ 7.1 kpc from the Galactic center 
(Table 9).The current cluster parameters exhibit a rather poor agreement with 
those derived by Tadross (2008) from 2MASS data. As in NGC\,5381, our 
larger E(B-V) value compared to that of Tadross (2008) may partly explain 
our lower age estimate.

\section{Conclusions}
We have presented new CCD Washington $CT_1$ photometry in the field of 
four Galactic OCs projected onto the two inner quadrants of the 
Galactic plane. These data were obtained with the main purpose of 
estimating the cluster fundamental astrophysical parameters. We 
performed a star count analysis of the cluster fields to assess the 
clusters' reality as over-densities of stars with respect to the 
field and estimated the cluster radii. We determined the center of 
the clusters by finding the maximum surface number density of the 
stars in each cluster. New equatorial coordinates for the 2000.0 
epoch are now provided. We outlined possible solutions for cluster 
fundamental parameters by matching theoretical isochrones, which 
reasonably reproduce the main cluster features in the $(C-T_1,T_1)$ 
CMDs. In all cases, the best fits were obtained using solar 
metallicity isochrones. BH\,211 was found to be the oldest 
object of our sample with an age of around 1.0 Gyr. Czernik\,37, 
the most heavily reddened cluster of the sample, with a mean colour excess 
$E(B-V)$ = 1.47, is very likely affected by differential reddening. 
BH\,84 is located 
in the fourth Galactic quadrant just before the tangent to the Carina 
branch of the Carina-Sagittarius spiral arm. This cluster turned 
out to be much older than previously believed. Conversely, NGC\,5381 
was found to be much younger than previously reported. It appears to 
have a relatively small but conspicuous nucleus and a low-density extended 
corona. We estimated the angular core and corona radii as $\sim$ 2.2' and 
$\sim$ 5.7', respectively. The derived fundamental properties for the studied 
clusters are listed in Table 9. Previous estimates of cluster parameters 
are listed in Table 10, for easy comparison with the present results. Since 
two of the studied clusters, BH\,211 and Czernik\,37, are in the VISTA Variables in the Via 
Lactea (VVV) survey \citep{mietal10}, additional observational information 
about these two objects can be found in this database.

\begin{table}
\caption{Basic parameters of the four open clusters}
\vspace{0.1cm}
\begin{tabular}{@{}|llcccc|lc|}\hline
\multicolumn{6}{|c|}{WEBDA} &\multicolumn{2}{c|}{This study}  \\
Cluster  & $\alpha$$_{\rm 2000}$ &  $\delta$$_{\rm 2000}$ & {\it l} & $b$ & Diam. & $\alpha$$_{\rm 2000}$ & $\delta$$_{\rm 2000}$ \\
         & (h m s)             & ($^\circ$ ' ")  & ($^\circ$)  & ($^\circ$) & (') & (h m s) & ($^\circ$ ' ") \\
\hline
\hline
BH\,84         & 10 01 19 & -58 13 00  & 280.06  & -2.42  & 4.5  & 10 01 19  & -58 13 33 \\
NGC\,5381      & 14 00 41 & -59 35 12  & 311.60  &  2.11  & 11.0 & 14 00 41  & -59 35 20 \\
BH\,211        & 17 02 11 & -41 06 00  & 344.97  &  0.46  & 4.0  & 17 02 10  & -41 05 57 \\
Czernik\,37    & 17 53 16 & -27 22 00  &   2.22  & -0.64  & 3.0  & 17 53 17  & -27 22 36 \\
\hline
\end{tabular}
\end{table}
         
\begin{center}
\begin{longtable}{|ccccc|}
\caption{Observation log of observed clusters} \\
\hline
Cluster    &   Date   & Filter   &  Exposure  & Airmass  \\
           &          &          &  (sec)     &  (")  \\
\hline
\hline
\endfirsthead
\multicolumn{5}{c}
{\tablename\ \thetable\ -- \textit{continued}} \\
\hline
Cluster    &   Date   & Filter    &  Exposure  & Airmass  \\
          &           &           & (sec)      & (")  \\
\hline
\hline
\endhead
\hline 
\endfoot
\hline
\endlastfoot
BH\,84   &  May 9, 2008  &  $C$   &  30  &  1.13  \\
         &               &  $C$   &  45  &  1.13  \\
         &               &  $C$   & 300  &  1.13  \\
         &               &  $C$   & 450  &  1.13  \\
         &               &  $R$   &   5  &  1.13  \\
         &               &  $R$   &   7  &  1.12  \\
         &               &  $R$   &  30  &  1.12  \\
         &               &  $R$   &  45  &  1.12  \\
\hline
NGC\,5381 &  May 11, 2008  &  $C$  &   90  &  1.15  \\
          &                &  $C$  &  120  &  1.15  \\
          &                &  $C$  &  600  &  1.15  \\
          &                &  $C$  &   30  &  1.16  \\
          &                &  $R$  &    3  &  1.16  \\
          &                &  $R$  &  120  &  1.16  \\
          &                &  $R$  &  120  &  1.16  \\
\hline
BH\,211  &                 &  $C$  &   30  &  1.02  \\
         &                 &  $C$  &   45  &  1.02  \\
         &                 &  $C$  &  300  &  1.02  \\
         &                 &  $C$  &  450  &  1.02  \\
         &                 &  $R$  &    5  &  1.02  \\
         &                 &  $R$  &    7  &  1.02  \\
         &                 &  $R$  &   30  &  1.02  \\
         &                 &  $R$  &   45  &  1.02  \\
\hline
Czernik\,37  &  May 10, 2008  &  $C$  &  30 &  1.00 \\
             &                &  $C$  &  45 &  1.00  \\
             &                &  $C$  & 300 &  1.00  \\
             &                &  $C$  & 450 &  1.00  \\
             &                &  $R$  &  30 &  1.00  \\
             &                &  $R$  &  45 &  1.00  \\
            &                &  $R$  &   5 &  1.00  \\
             &                &  $R$  &   7 &  1.00  \\  
\end{longtable}
\end{center}

\begin{table}[h]
\begin{center}
\caption{Standard system mean calibration coefficients}
\vspace{0.1cm}
\begin{tabular}{@{}|cc|}\hline
$C$     &   $T_1$   \\
\hline
\hline
$a_1$ = 3.61 $\pm$ 0.03   &  $b_1$ = 3.04 $\pm$ 0.02  \\
$a_2$ = 0.56 $\pm$ 0.01   &  $b_2$ = 0.33 $\pm$ 0.02  \\
$a_3$ = -0.19 $\pm$ 0.01 & $b_3$ = -0.03 $\pm$ 0.01 \\
\hline
\end{tabular}
\end{center}
\end{table}

\begin{table}
\caption[]{CCD $CT_1$ data of stars in the field of BH\,84}
\vspace{0.1cm}
\begin{tabular}{|cccccccc|}
\hline
Star & $X$\hspace{0.2cm} &  $Y$\hspace{0.2cm} & $T_{1}$  & $\sigma$$T_1$  & $C-T_1$  & $\sigma$$(C-T_1)$ & n \\
     & (pixel)           & (pixel)           &  (mag)  & (mag)        &  (mag)   & (mag)           &  \\
\hline
495 & 1870.105 & 584.378 & 17.590 & 0.094 & 3.237 & 0.071 & 1 \\
496 &  926.421 & 585.257 & 15.910 & 0.018 & 1.955 & 0.015 & 2 \\
497 & 1568.217 & 585.395 & 13.705 & 0.010 & 1.653 & 0.009 & 1 \\
-   &    -              &    -              &    -    &   -          &    -     &   -             & -  \\
 -   &    -              &    -              &    -    &   -          &    -     &   -             & -   \\
\hline
\end{tabular}
\end{table}

\begin{table}
\caption[]{CCD $CT_1$ data of stars in the field of NGC\,5381}
\vspace{0.1cm}
\begin{tabular}{|cccccccc|}
\hline
Star  &  $X$\hspace{0.2cm}  & $Y$\hspace{0.2cm} &  $T_{1}$  &  $\sigma$$T_1$  &  $C-T_1$  &  $\sigma$$(C-T_1)$  &  n  \\
      &  (pixel)            &  (pixel)          &  (mag)    &  (mag)          &  (mag)    &  (mag)              &     \\
\hline
499 & 918.737 & 415.590 & 13.625 & 0.005 & 1.874 & 0.005 & 2 \\
500 & 501.666 & 418.198 & 14.553 & 0.005 & 1.225 & 0.005 & 1 \\
501 & 1000.223& 422.144 & 16.461 & 0.022 & 1.801 & 0.017 & 2 \\
-    &   -                 &     -             &     -     &    -    &    -    &    -   &  -   \\
 -    &   -                 &     -             &     -     &    -    &    -    &    -   &  -   \\
\hline
\end{tabular}
\end{table}

\begin{table}
\caption[]{CCD $CT_1$ data of stars in the field of BH\,211}
\vspace{0.1cm}
\begin{tabular}{|cccccccc|}
\hline
Star  &  $X$\hspace{0.2cm}  & $Y$\hspace{0.2cm}  &  $T_{1}$  & $\sigma$$T_1$  & $C-T_1$  &  $\sigma$$(C-T_1)$  &  n  \\
      &   (pixel)           &  (pixel)           &  (mag)    & (mag)          & (mag)    & (mag)             &     \\
\hline
496 & 1255.054 & 1053.165 & 16.858 & 0.043 & 2.626 & 0.034 & 2 \\
497 & 1137.926 & 2039.064 & 16.899 & 0.052 & 2.597 & 0.040 & 1 \\
498 &  983.109 & 1055.830 & 15.363 & 0.016 & 2.018 & 0.014 & 1 \\
  -   &    -      &   -      &    -    &   -    &  -     &   -     &  -  \\
  -   &    -      &   -      &    -    &   -    &  -     &   -     &  -  \\
\hline
\end{tabular}
\end{table}

\begin{table}
\caption[]{CCD $CT_1$ data of stars in the field of Czernik\,37}
\vspace{0.1cm}
\begin{tabular}{|cccccccc|}
\hline
Star  & $X$\hspace{0.2cm}  & $Y$\hspace{0.2cm}  &  $T_{1}$  &  $\sigma$$T_1$  & $C-T_1$  & $\sigma$$(C-T_1)$  &  n  \\
      &  (pixel)           &  (pixel)           &  (mag)    &  (mag)          & (mag)    & (mag)            &    \\
\hline
500 & 1480.077 & 796.860 & 15.898 & 0.038 & 3.126 & 0.033 & 1 \\
501 &  961.983 & 801.728 & 15.552 & 0.021 & 2.819 & 0.018 & 2 \\
502 & 1114.297 & 805.020 & 15.474 & 0.041 & 2.794 & 0.03 & 1 \\
-     &   -       &   -     &   -       &   -     &   -    &   -      &  -  \\
  -   &   -       &   -      &   -      &   -    &   -     &   -      &   -  \\
\hline
\end{tabular}
\end{table}
 
\begin{table}
\caption{Cluster sizes}
\vspace{0.1cm}
\begin{tabular}{|l|cc|cc|cc|}
\hline
Cluster      & \multicolumn{2}{|c|}{$r_{FWHM}$}      &  \multicolumn{2}{c|}{$r_{clean}$}  & \multicolumn{2}{c|}{$r_{cl}$}  \\
             & (px) &  (pc)     &   (px)  &  (pc)   &   (pix)  & (pc)  \\
\hline
BH\,84       & 150  &  1.0      &   200   &  1.3    &   550    &  3.6  \\
NGC\,5381    & 150  &  0.8      &   280   &  1.4    &  1200    &  6.1  \\
BH\,211      & 150  &  0.4      &   250   &  0.7    &   580    &  1.6  \\
Czernick\,37 & 220  &  0.6      &   320   &  0.9    &   550    &  1.5  \\
\hline
\end{tabular}
\end{table}
                
\begin{table}
\caption{Fundamental properties of the observed clusters}
\vspace{0.1cm}
\begin{tabular}{|lcccccccc|}
\hline
  Cluster  &  $E(B-V)$  &  $d$     &     Age  &  [Fe/H]  &   $X$    &  $Y$    &  $Z$    & $R_{GC}$   \\
           &   (mag)    &  (kpc)   &    (Myr) &   (dex)   &   (kpc)  &  (kpc)  &  (kpc)  &  (kpc)      \\
\hline
\hline
BH\,84      &   0.63 $\pm$ 0.05  & 3.37 $\pm$ 0.48  & 560$^{+150}_{-110}$  & 0.0  &  7.91  &  -3.32  &  -0.14  &  8.58  \\
NGC\,5381  &   0.46 $\pm$ 0.04  & 2.63 $\pm$ 0.40  & 250$^{+65}_{-50}$   & 0.0  &  6.76  &  -1.97  &   0.10  &  7.04  \\
BH\,211    &   0.61 $\pm$ 0.05  & 1.44 $\pm$ 0.21  & 1000$^{+260}_{-200}$ & 0.0  &  7.11  &  -0.37  &   0.01  &  7.12 \\
Czernik\,37 &  1.47 $\pm$ 0.25  & 1.44 $\pm$ 0.86  & 250$^{+100}_{-65}$   & 0.0  &  7.06  &   0.06  &  -0.02  &  7.06  \\
\hline
\end{tabular}
\end{table}   

\begin{table}
\caption{Previous stimates of cluster parameters}
\vspace{0.1cm}
\begin{tabular}{|lcccc|}
\hline
  Cluster  &  $E(B-V)$  &  $d$     &     Age  &  Reference \\
           &   (mag)    &  (kpc)   &    (Myr) &   \#   \\
\hline
\hline
BH\,84      &   0.60   & 2.92 $\pm$ 0.19  & 18    & 1  \\
NGC\,5381   &   0.06   & 1.17 $\pm$ 0.05  & 1600  & 2  \\
BH\,211     &   0.48   & 1.38 $\pm$ 0.09  & 1600  & 1  \\
Czernik\,37 &   1.03   & 1.73 $\pm$ 0.08  & 600   & 3  \\
\hline
\end{tabular}

References: (1) Bukowiecki et al.(2011); (2) Tadross (2011); (3) Tadross (2008)

\end{table}

\begin{figure}
\begin{center}
\includegraphics*[width=12cm]{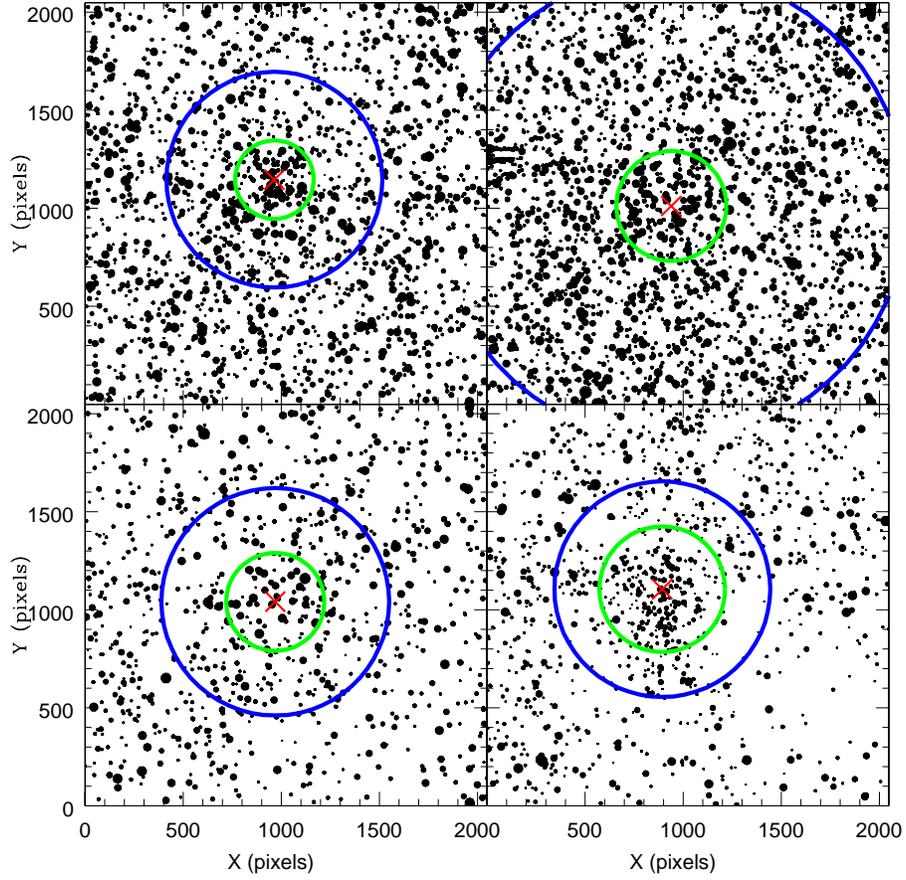}
\end{center}
\caption{Schematic finding charts of the stars observed in BH\,84 (top left), NGC\,5381 (top right),
BH\,211 (bottom left) and Czernik\,37 (bottom right). North is up and East is to the left. 
The sizes of the plotting symbols are proportional to the $T_1$ brightness of the stars. Two circles 
$r_{clean}$ and $r_{cl}$ wide are shown around the cluster centers ({\it crosses}).} 
\label{fig1} 
\end{figure}

\begin{figure}
\begin{center}
\includegraphics*[width=12cm]{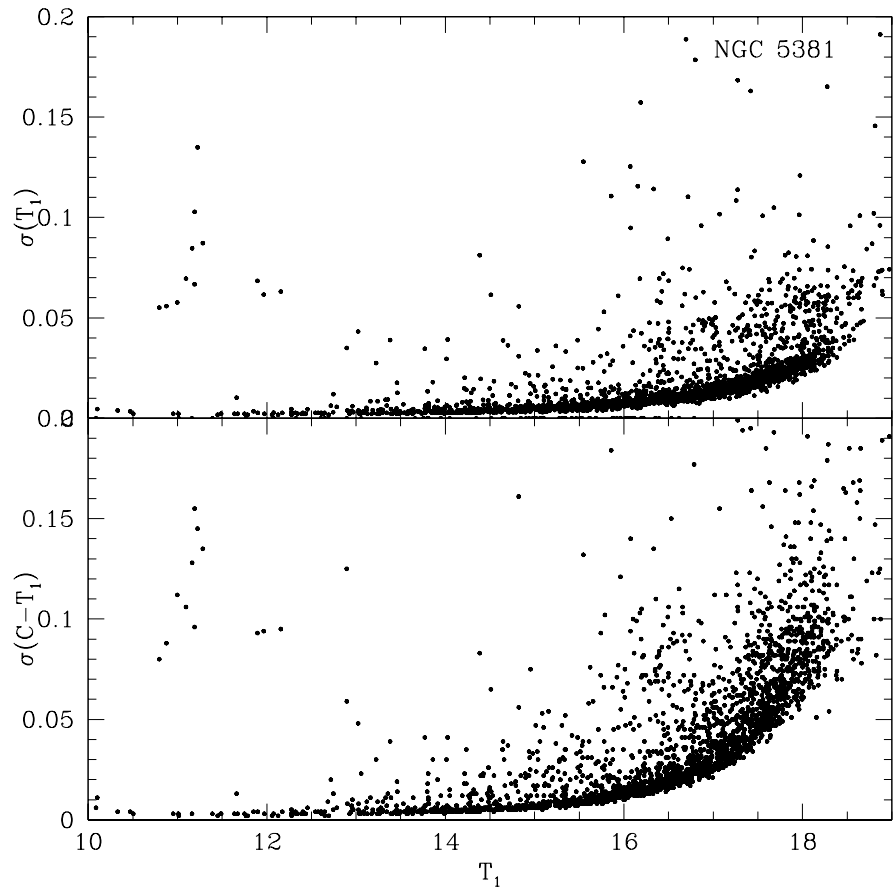}
\end{center}
\caption{$T_1$ magnitude and C-T$_1$ color photometric errors as a function of $T_1$ for 
stars measured in the field of NGC\,5381.}
\label{fig2}
\end{figure}

\begin{figure}
\begin{center}
\includegraphics*[width=12cm]{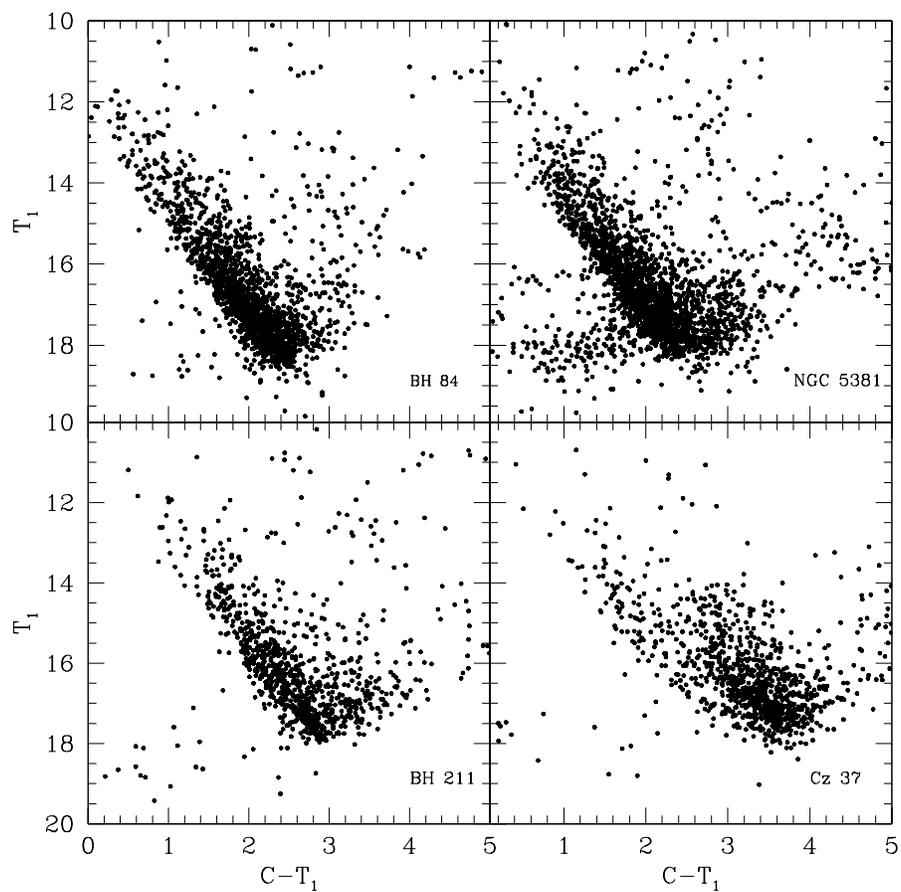}
\end{center}
\caption{($C-T_1$,$T_1)$ CMDs for stars observed in the field of BH\,84, NGC\,5381, BH\,211 and Czernik\,37.}
\label{fig3}
\end{figure}

\begin{figure}
\begin{center}
\includegraphics*[width=12cm]{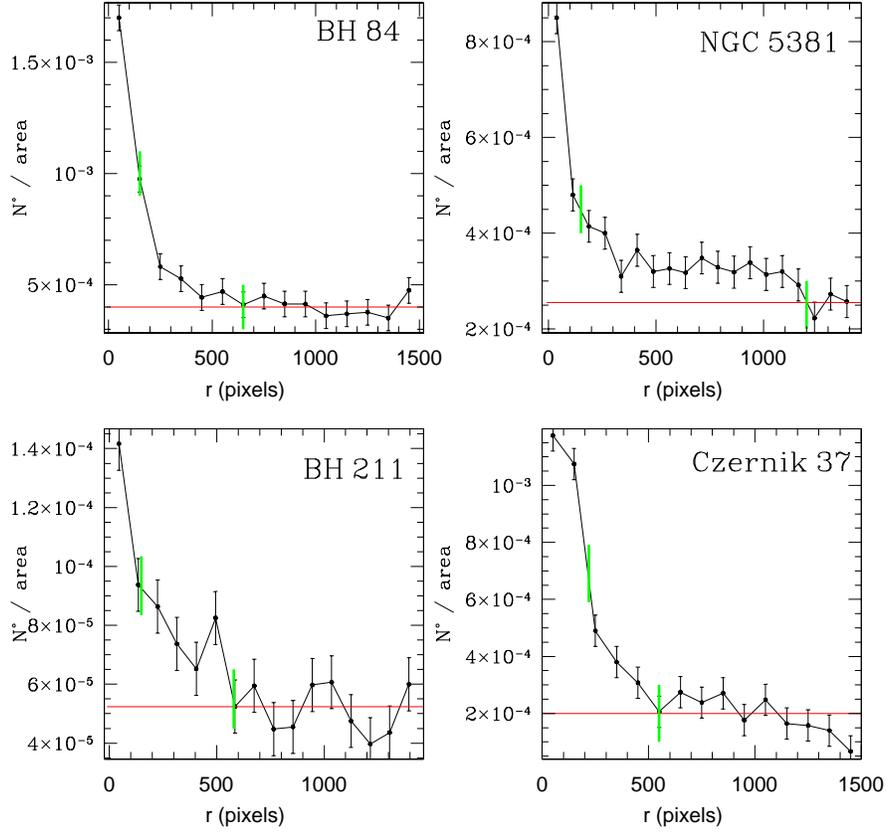}
\end{center}
\caption{Cluster stellar density radial profiles normalized to a circular area of 50 pixel radius. The
radius at the FWHM (r$_{FWHM}$) and the adopted cluster radius (r$_{cl}$) are indicated by  
green vertical lines. The red horizontal lines represent the measured background levels.}
\label{fig4}
\end{figure}

\begin{figure}
\begin{center}
\includegraphics*[width=12cm]{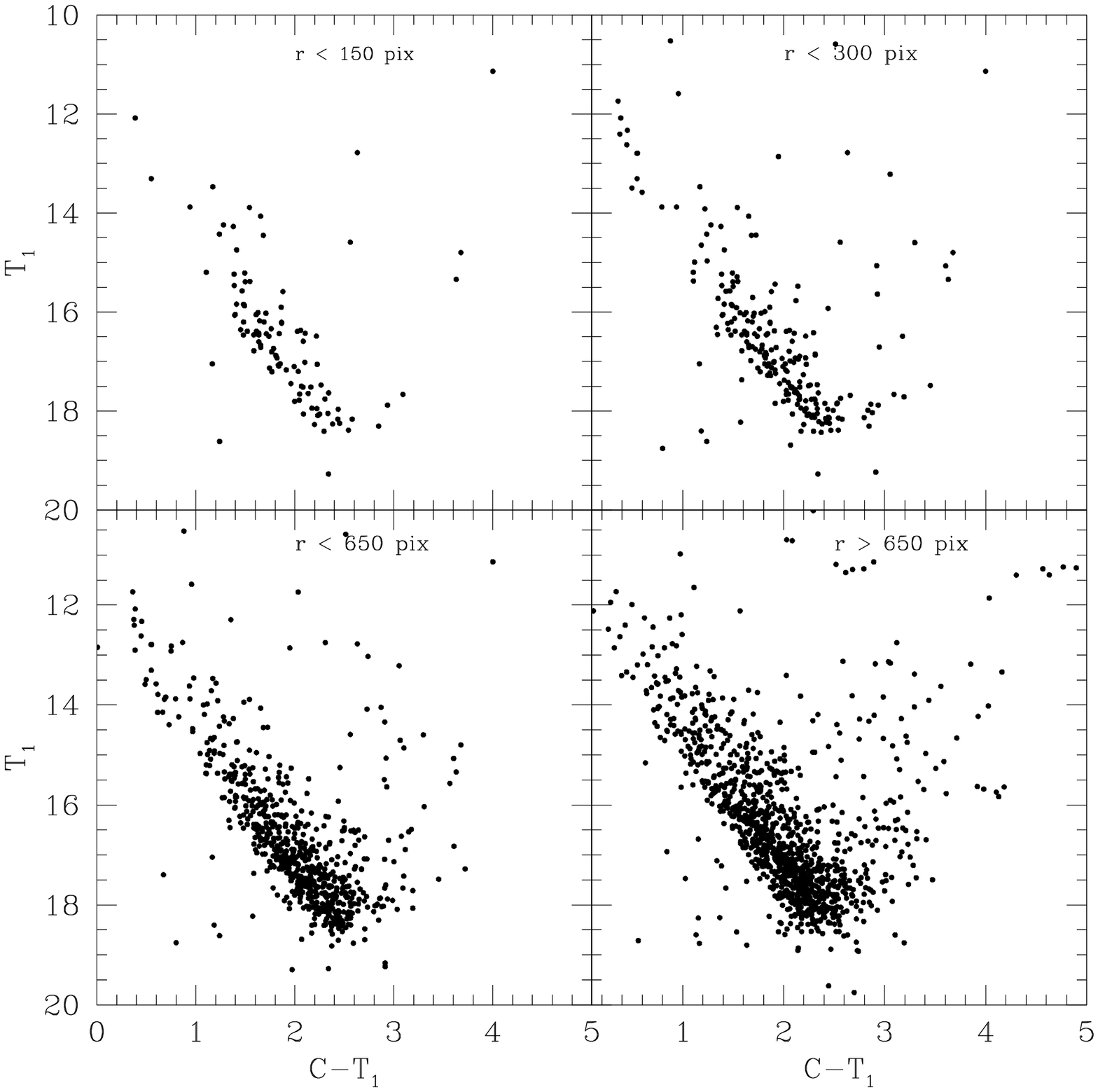}
\end{center}
\caption{CMDs for stars observed in different extracted circular regions around BH\,84 center 
as indicated in each panel.}
\label{fig5}
\end{figure}

\begin{figure}
\begin{center}
\includegraphics*[width=12cm]{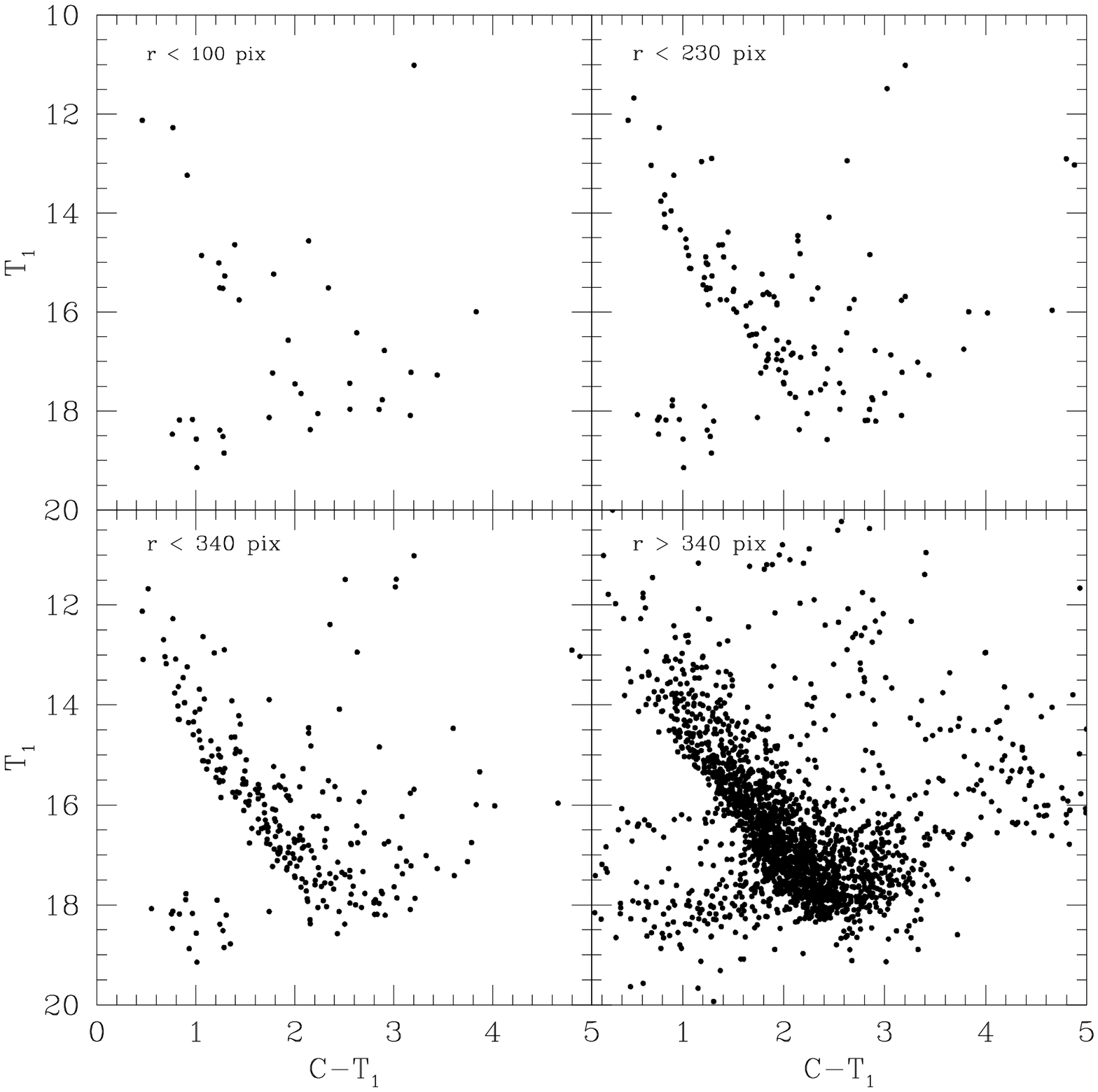}
\end{center}
\caption{CMDs for stars observed in different extracted circular regions around NGC\,5381 center  
as indicated in each panel.}
\label{fig6}
\end{figure}

\begin{figure}
\begin{center}
\includegraphics*[width=12cm]{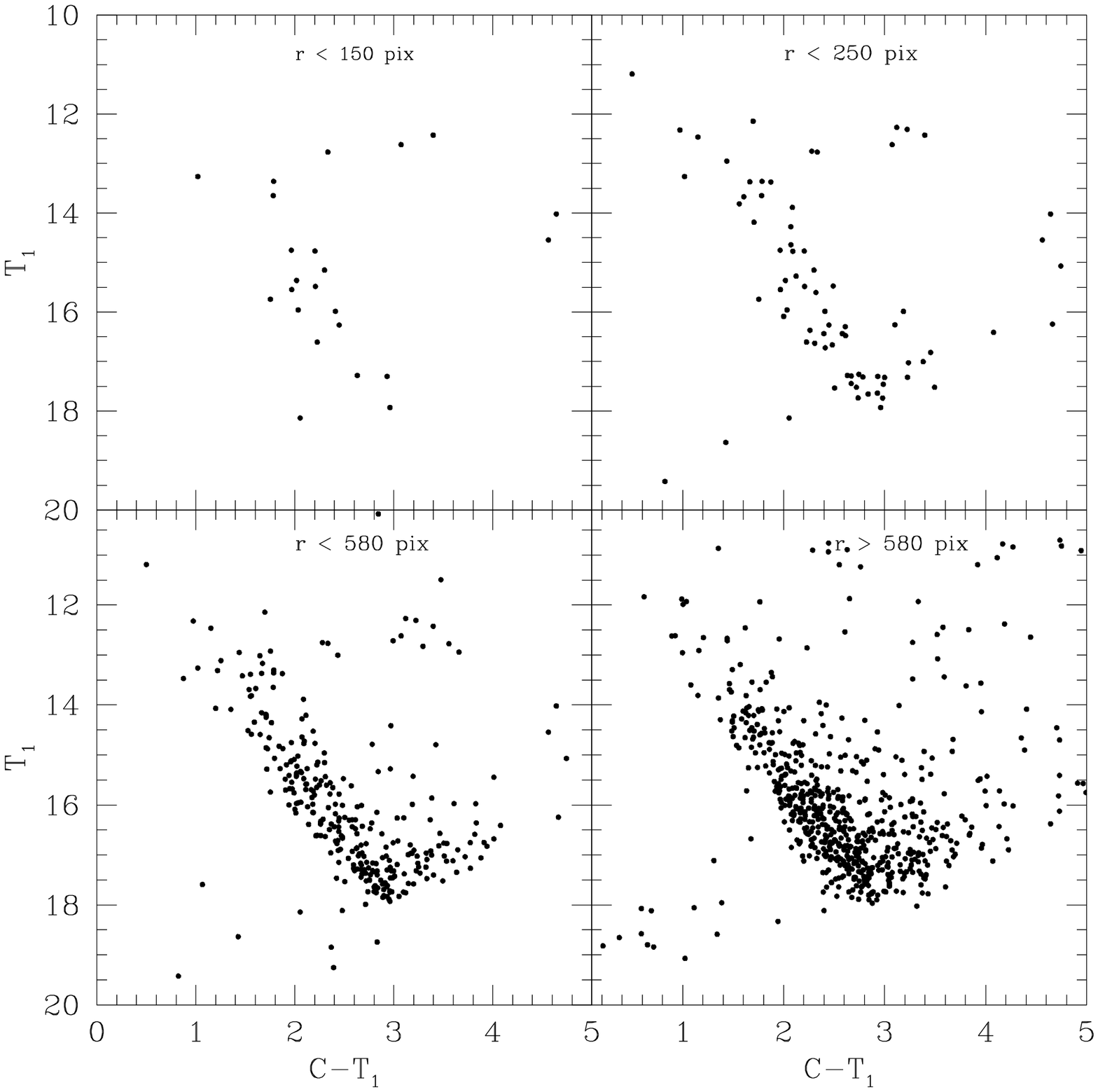}
\end{center}
\caption{CMDs for stars observed in different extracted circular regions around BH\,211 center  
as indicated in each panel.}
\label{fig7}
\end{figure}

\begin{figure}
\begin{center}
\includegraphics*[width=12cm]{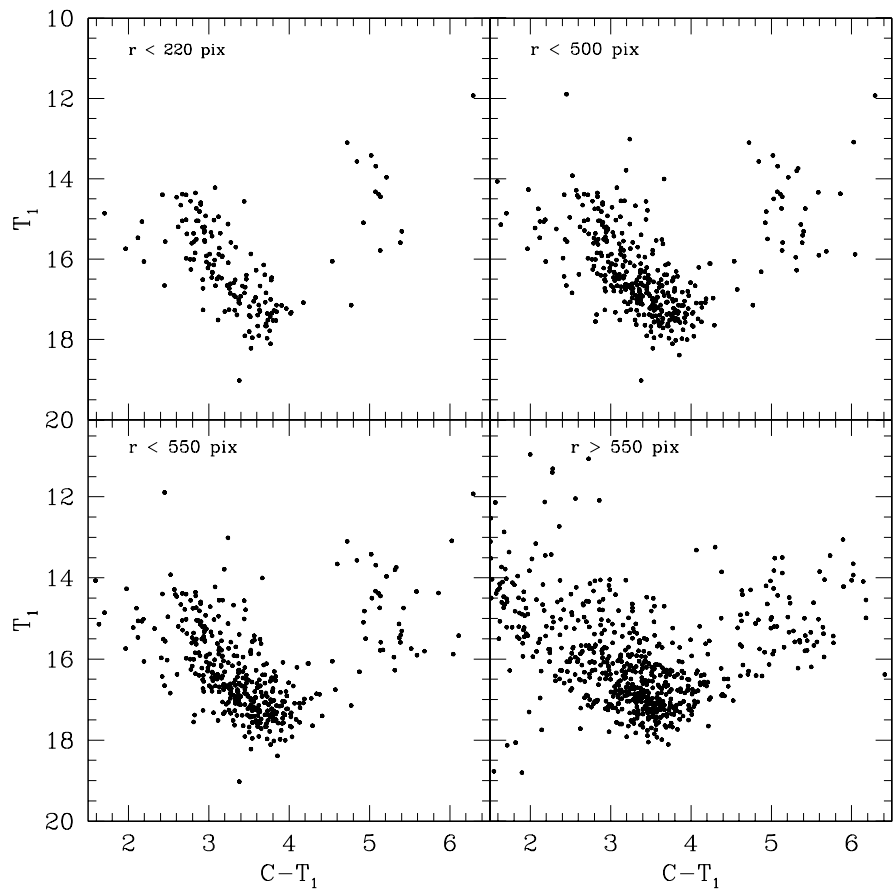}
\end{center}
\caption{CMDs for stars observed in different extracted circular regions around Czernik\,37 center  
as indicated in each panel.}
\label{fig8}
\end{figure}

\begin{figure}
\begin{center}
\includegraphics*[width=12cm]{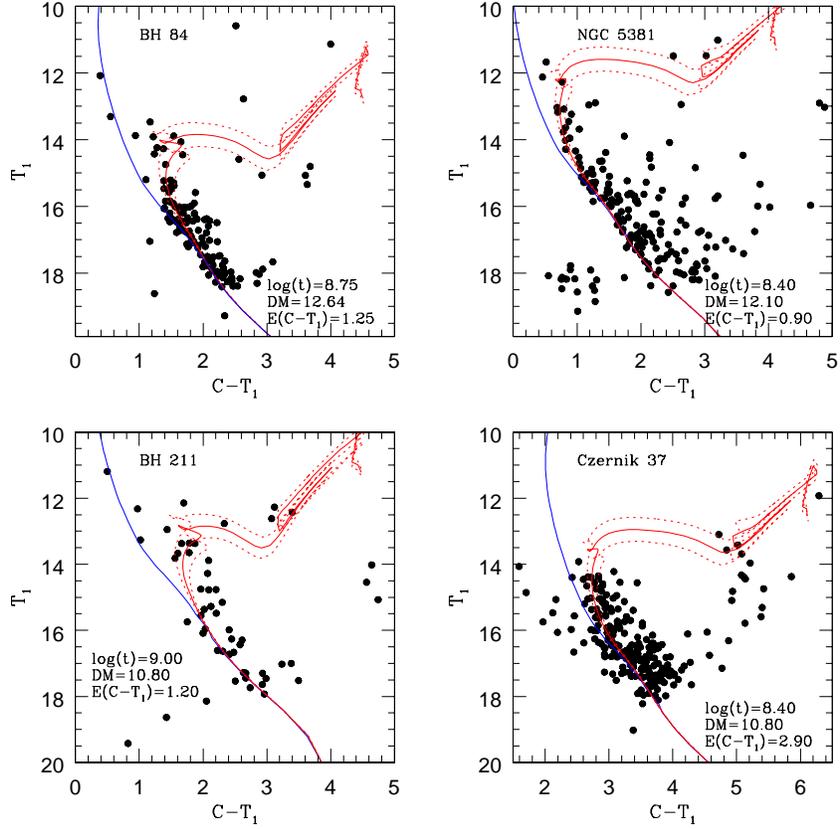}
\end{center}
\caption{r $<$ r$_{clean}$ (C-$T_1$,$T_1$) CMDs for stars in: BH\,84 (top left), NGC\,5381 (top right), BH\,211 
(bottom left) and Czernik\,37 (bottom right). The ZAMS and the adopted isochrones from Girardi et al. 
(2002) are overplotted with solid lines. The isochrones associated to the cluster age errors are indicated by 
dashed lines, for comparison purposes.}
\end{figure}

\section{Acknowledgements}

J.J. Clari\'a, T. Palma and A.V. Ahumada are gratefully indebted to the CTIO staff for their hospitality and support
during the observing run. We also thank the anonymous referee for his/her valuable comments and 
suggestions. This research was partially supported by the Argentinian institutions CONICET, SECYT 
(Universidad Nacional de C\'ordoba) and Agencia Nacional de Promoci\'on Cient\'{\i}fica y Tecnol\'ogica 
(ANPCyT). We have used both the SIMBAD database, operated at CDS, Strasbourg, France, and the NASA's 
Astrophysics Data System. This work is based on observations made at Cerro Tololo Inter-American 
Observatory, which is operated by AURA, Inc., under cooperative agreement with the NSF.

\bibliographystyle{elsarticle-harv}

\end{document}